\RequirePackage{fix-cm}
\documentclass[smallextended]{svjour3}       
\smartqed  
\usepackage{graphicx}
\usepackage{amsmath}

\usepackage{subfigure}
\journalname{Experimental Astronomy}
\begin{document}

\title{Scaling Radio Astronomy Signal Correlation on Heterogeneous Supercomputers Using Various Data Distribution Methodologies}
\titlerunning{ }


\author{Ruonan Wang         \and
        Christopher Harris 
}


\institute{Ruonan Wang \at
              ICRAR, M468, The University of Western Australia, 35 Stirling Highway,\\
 Crawley, WA 6009, Australia \\
              \email{jason.wang@icrar.org}           
           \and
           Christopher Harris \at
	      iVEC@UWA, M024, The University of Western Australia, 35 Stirling Highway,\\
 Crawley, WA 6009, Australia\\
              \email{christopher.harris@uwa.edu.au}           
}

\date{Received: 26 Nov 2012 / Accepted: 05 May 2013}

\maketitle

\begin{abstract}
Next generation radio telescopes will require orders of magnitude more computing power to provide a view of the universe with greater sensitivity. In the initial stages of the signal processing flow of a radio telescope, signal correlation is one of the largest challenges in terms of handling huge data throughput and intensive computations. We implemented a GPU cluster based software correlator with various data distribution models and give a systematic comparison based on testing results obtained using the Fornax supercomputer. By analyzing the scalability and throughput of each model, optimal approaches are identified across a wide range of problem sizes, covering the scale of next generation telescopes. 

\end{abstract}

\section{Introduction}
\label{intro}

Signal correlation is one of the most computationally demanding and communication intensive tasks in the signal processing flow of a radio telescope array. It has been traditionally processed using field programmable gate arrays (FPGAs) to achieve excellent power efficiency. However, high development challenges and lack of portability make it an expensive task either to design a system from scratch, to scale an existing one or to introduce new functionalities. With the fast development of general purpose hardware platforms, it is likely that at some point the relatively low development cost and high flexibility of software correlators make them a viable option.

There have been a growing number of software correlator projects over the last decade. The most widely used CPU cluster based VLBI correlator DiFX designed by Deller et al. \cite{DiFX,DiFX2} implemented a time division multiplexed system, in which the inter-node synchronization is less critical and hence it achieves excellent performance even given unbalanced computing resources or non-ideal network conditions. Recent research by Dodson et al. \cite{richard} showed that DiFX can be efficiently implemented on supercomputers with Infiniband networks as well as on the Intel MIC architecture, and it scales linearly up to 50 nodes after which network bottlenecks cut in. 

Another well-known software correlator was designed for the Low Frequency Array (LOFAR). Being one of the first new generation telescopes intensively using interferometry techniques, LOFAR was also one of the first real world projects to use a dedicated software correlator. A Blue Gene/L supercomputer is used in the LOFAR system for correlation and post-correlation processing by Romein et al. \cite{lofar,lofarcorr}. Computationally intensive jobs in the LOFAR software system are optimized using assembly language and as a result, it has achieved 98\% of the peak floating-point capability of the hardware architecture. 

While the CPU-based software correlators proved the capability, GPUs (Graphic Processing Unit) appear to be increasingly applicable for this type of work. In a comparison of correlation on different hardware architectures by Nieuwpoort et al. \cite{IBM}, NVIDIA GPUs showed the best absolute performance and the second best power efficiency, which revealed the feasibility of building a powerful GPU-based correlation system. GPUs were first used for correlation a decade ago by Schaaf et al. \cite{cots}, when graphic programming techniques such as the Cg language had to be heavily involved to get a general computing problem solved on a GPU. 

Over the last few years there have been several GPU-based software correlators. These took advantage of NVIDIA's Compute Unified Device Architecture (CUDA), in which GPUs can be treated as generic computing devices in addition to graphic chips. This significantly reduced the programming challenge, and hence more efforts could be put into the optimization, rather than making algorithms compatible with the hardware. The first CUDA-based GPU correlator designed by Harris et al. \cite{chris} took advantage of the CUFFT library for its F-engine and implemented a series of X-engines in different parallel fashions, which achieved a considerable performance gain compared with CPU correlators. Another project conducted by Wayth et al. \cite{mwa} implemented similar parallel approaches to those presented by Harris et al. \cite{chris} and constructed a real-time correlator for the prototype of the Murchison Widefield Array (MWA). The most recent work by Clark et al. \cite{xGPU} presented a highly optimized implementation on NVIDIA's GTX480 GPU and achieved 79\% of the peak single precision capacity of the hardware architecture. 

The world's largest radio telescope, the Square Kilometer Array (SKA), is also considering a correlation system based on GPU clusters as described by D'Addario \cite{SKA}. However, previous research has focused on single-GPU approaches with very little consideration given to data distribution across multiple GPUs. The data distribution pattern used in CPU cluster correlators are yet to be verified with GPU clusters given the number of distinctive features of GPU correlator engines. 

This paper presents a software correlator for heterogeneous high performance computing clusters, especially GPU clusters, mainly focusing on data distribution models. Two space-division network models are proposed in this paper and are compared with a re-implemented time-division model which was first introduced by Deller et al. \cite{DiFX}. The correlator engines presented by Harris et al. \cite{chris} are adopted and re-implemented in the Open Computing Language (OpenCL) for compatibility with different computing devices. The scope of this work is to investigate possible solutions for solving large-scale correlation problems such as those SKA would face. 

\section{FX Correlator}

There are two main approaches to radio astronomy signal correlation. The first, a lag or XF correlator, correlates signals in the time domain, before transformation to the frequency domain via the Fourier Transform. This method is often used in hardware implementations where the initial correlation can be performed at lower bit precision. The second, an FX correlator, instead transforms the signals using the Fourier Transform, and then performs the correlation via conjugate multiplication. This method is predominantly used in software correlators, as it requires fewer total operations. In both methods the results are usually then accumulated. As this work will utilize the FX correlator, a brief mathematical introduction follows.

For a discrete time signal $s[n]$ of N samples, with $n \in [0,N-1]$, the Discrete Fourier Transform is first applied to obtain the spectra $S[k]$ for frequencies $k \in [0,N-1]$ as shown in Equation \ref{eq:dft}:

\begin{equation}
\label{eq:dft}
S[k] = \sum_{n=0}^{N-1} s[n] e^{-j (2 \pi / N) k n}
\end{equation}

Then for $S_{a,i}[k]$ and $S_{a,j}[k]$, where $a$ is the index of the spectra over time, $i$ and $j$ are the index of each signal in the pair, the complex visibilities $C_{ij}[k]$ are obtained using Equation \ref{eq:dft_corr}:

\begin{equation}
\label{eq:dft_corr}
C_{ij}[k]=\sum_{a=0}^{A-1}S_{a,i}^\ast[k] S_{a,j}[k]
\end{equation}

In an FX correlator implementation, the two steps are usually named the F-engine and the X-engine. This work takes advantage of the Apple OpenCL FFT to implement the F-engine. For the X-engine, the 1xGxG model used by Harris et al. \cite{chris} is adopted and re-implemented in OpenCL with modifications to fit cluster models.

\section{Time-division Model}

\begin{figure}
\includegraphics[width=0.75\textwidth]{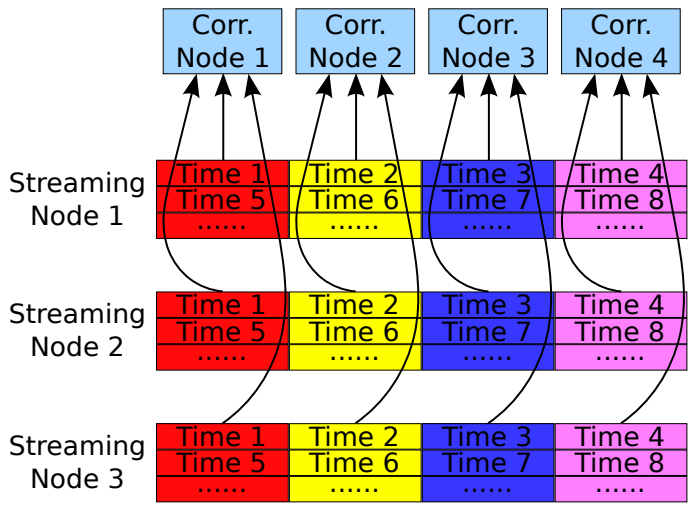}
\caption{Shown is the data distribution pattern of the time-division model. Each streaming node contains multiple input data streams. Data streams are divided into time slices and transferred separately to corresponding correlation nodes based on the time-division allocation. }
\label{fig:timedivision}    
\end{figure}

The time-division pattern for correlation, which was used by Deller et al. \cite{DiFX}, is the first data distribution model we implemented in this work. As shown in Figure \ref{fig:timedivision}, input data streams on streaming nodes are divided into time slices and distributed to correlation nodes. Each correlation node is responsible for some of the time slices across all input streams. An input stream here refers to the sampled digital data from an antenna, which does not include the case where the data is channelized into sub-bands or where multiple polarizations are present per stream. The time-division model is highly efficient in terms of data transfers as all input data chunks are transferred only once. Moreover, every correlation node processes independent data, and as a result, synchronization between correlation nodes becomes less important. 

However, as the time-division model was originally proposed for a CPU cluster correlator, simply replacing the FX engines with GPU implementations could potentially cause problems. Based on our preliminary testing, when the number of input data streams becomes very large, the efficiency of the FX engines drops dramatically. In this case the time-division model is not necessarily optimal on a GPU cluster even though it is highly efficient in terms of data transfers. Furthermore, the time-division model processes all baselines on a single node, thus when it comes to a point where the number of data streams is so large that the GPU memory is not able to hold all baseline data at a minimum length of a single FFT, the model would fail. Thus it is relevant to consider other data distribution models for GPU cluster correlators.

\section{Space-division Models}

\begin{figure}
\includegraphics[width=0.55\textwidth]{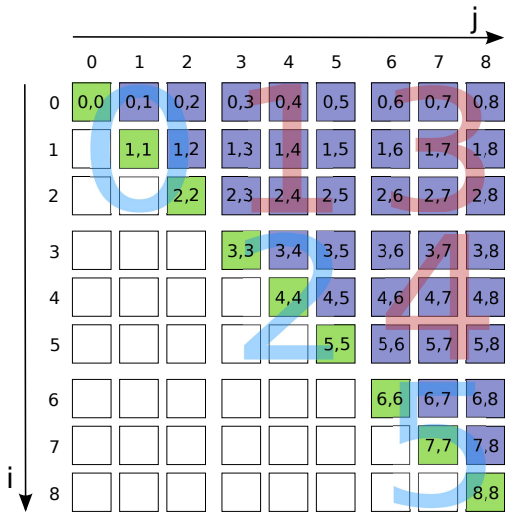}
\caption{Shown is the group pattern of correlation pairs. Each signal pair undergoing correlation is presented using a box, indexed by its constituent streams $i$ and $j$. Groups labeled using numbers in the larger font correspond to correlation nodes.}
\label{fig:group}    
\end{figure}

An alternative approach is to implement data distribution models based on division in space rather than time. Shown in Figure \ref{fig:group} are correlation jobs divided into groups based on the space-division pattern. Instead of processing all correlation pairs inside a single node and assigning different nodes with different time slices, a space-division model divides correlation pairs into groups, and each node assigned with a certain group is responsible for all time slices. Thus, given the total number of input streams, the number that a single node needs to process is reduced, which would improve the GPU X-engine performance for cases with a large number of streams. The exact number of streams per node would still be dependent on the total number of correlation nodes required to achieve real-time processing. Ultimately, to completely control the number of streams per node, a hybrid system would need to be used, but this is left for future investigations.

Space-division models involve necessary modifications to the X-engine, since X-engines designed for single GPU correlators process all input data streams at once in a triangle pattern for all non-redundant pairs, while some of the nodes in the space-division model need to process two parts of the input streams in a rectangle pattern for cross-correlations only. Moreover, it involves redundant data transfers as correlation nodes in the same row or column require the same input data. It is then of significant importance to design network topologies intelligent enough to handle the huge data efficiently. In this paper, we propose two network topologies to investigate the performance of the space-division model.

\subsection{Broadcasting Model}
\begin{figure}
\subfigure[Broadcasting Model]{
\includegraphics[width=0.5\textwidth]{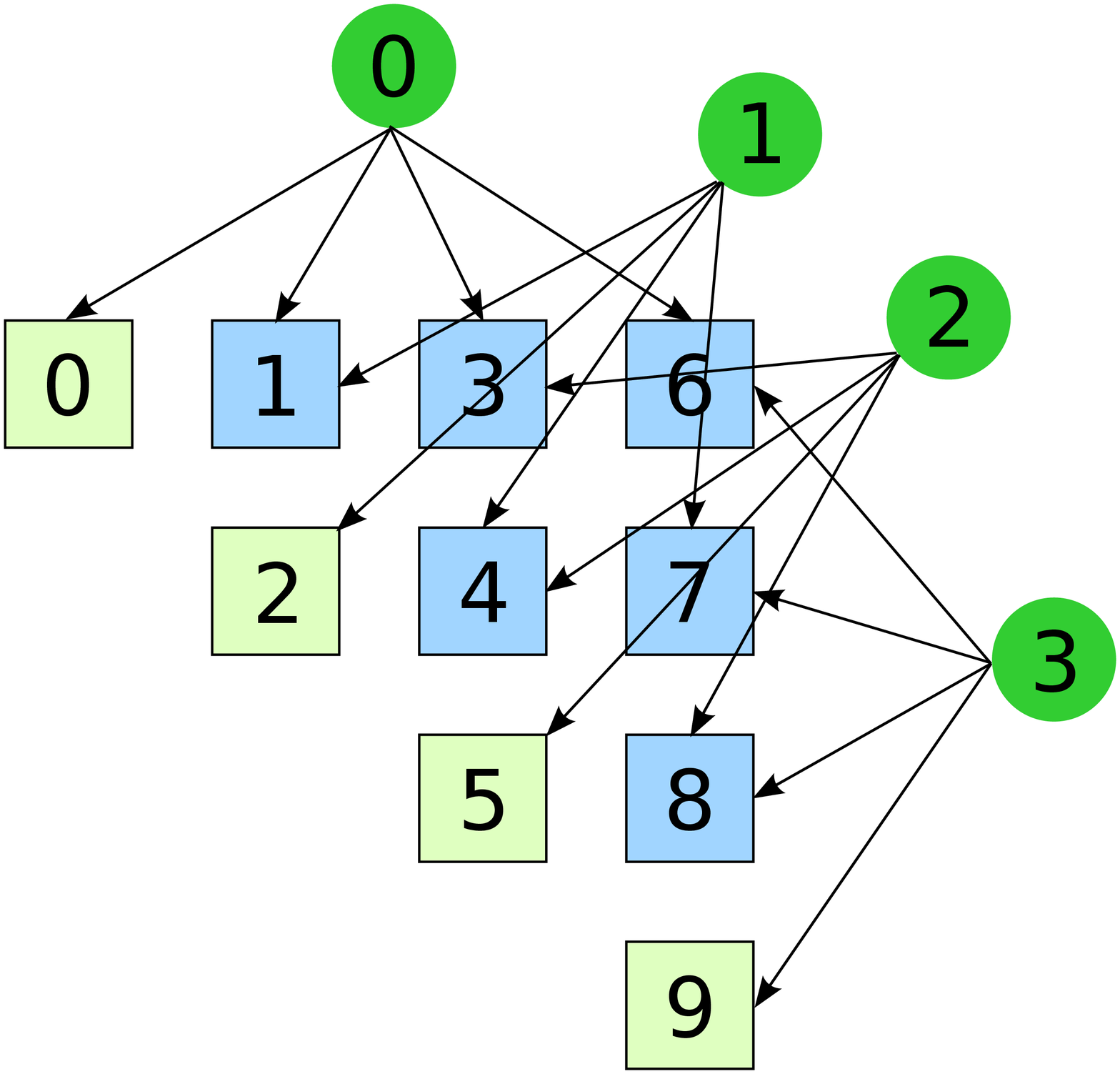}
\label{fig:broadcasting}    
}
\subfigure[Passing Model]{
\includegraphics[width=0.45\textwidth]{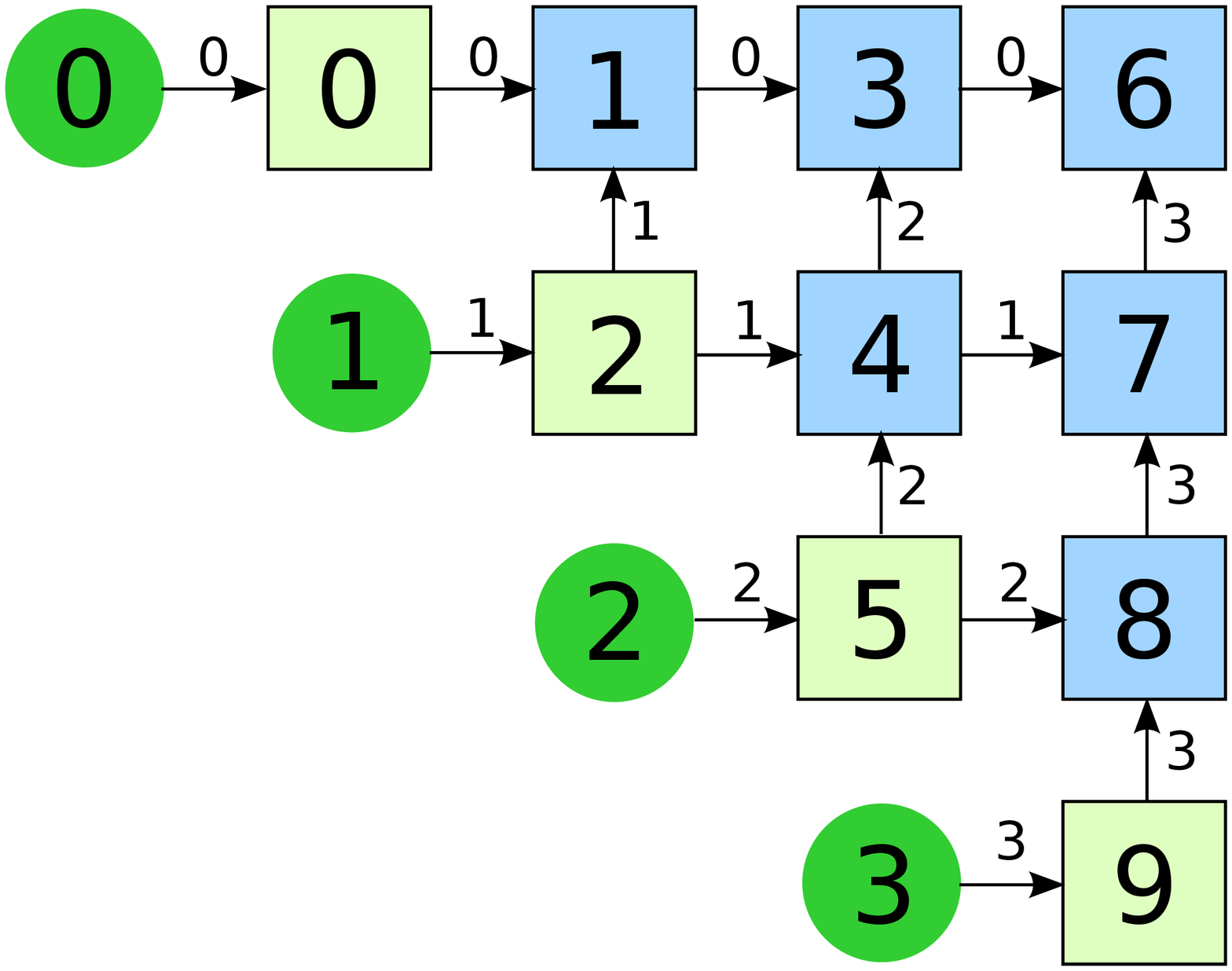}
\label{fig:passing}    
}
\caption{Shown are the topologies and data flows of our space-division network models. Circles represent data sources while squares are correlation nodes. Data sources can be only streaming nodes within the correlator allocation for the broadcasting model due to the broadcast and buffer management tasks involved, while in the passing model, they could also be external streaming sources or data files instead.}
\label{fig:spacemodels}
\end{figure}

The first space-division based network topology we designed is the broadcasting model shown in Figure \ref{fig:broadcasting}, which uses streaming nodes to broadcast the data across correlation nodes. Equation \ref{eq:oc_os} shows how the number of streaming nodes, $n_s$, varies with the number of correlation nodes, $n_c$.

\begin{align}
\label{eq:oc_os}
n_c &= \frac { n_s(n_s+1)}{2} \notag \\
n_s &= \frac {\sqrt{8n_c+1}-1}{2}
\end{align}

\begin{figure}
\subfigure[Broadcast (Multicast)]{
\includegraphics[width=0.5\textwidth]{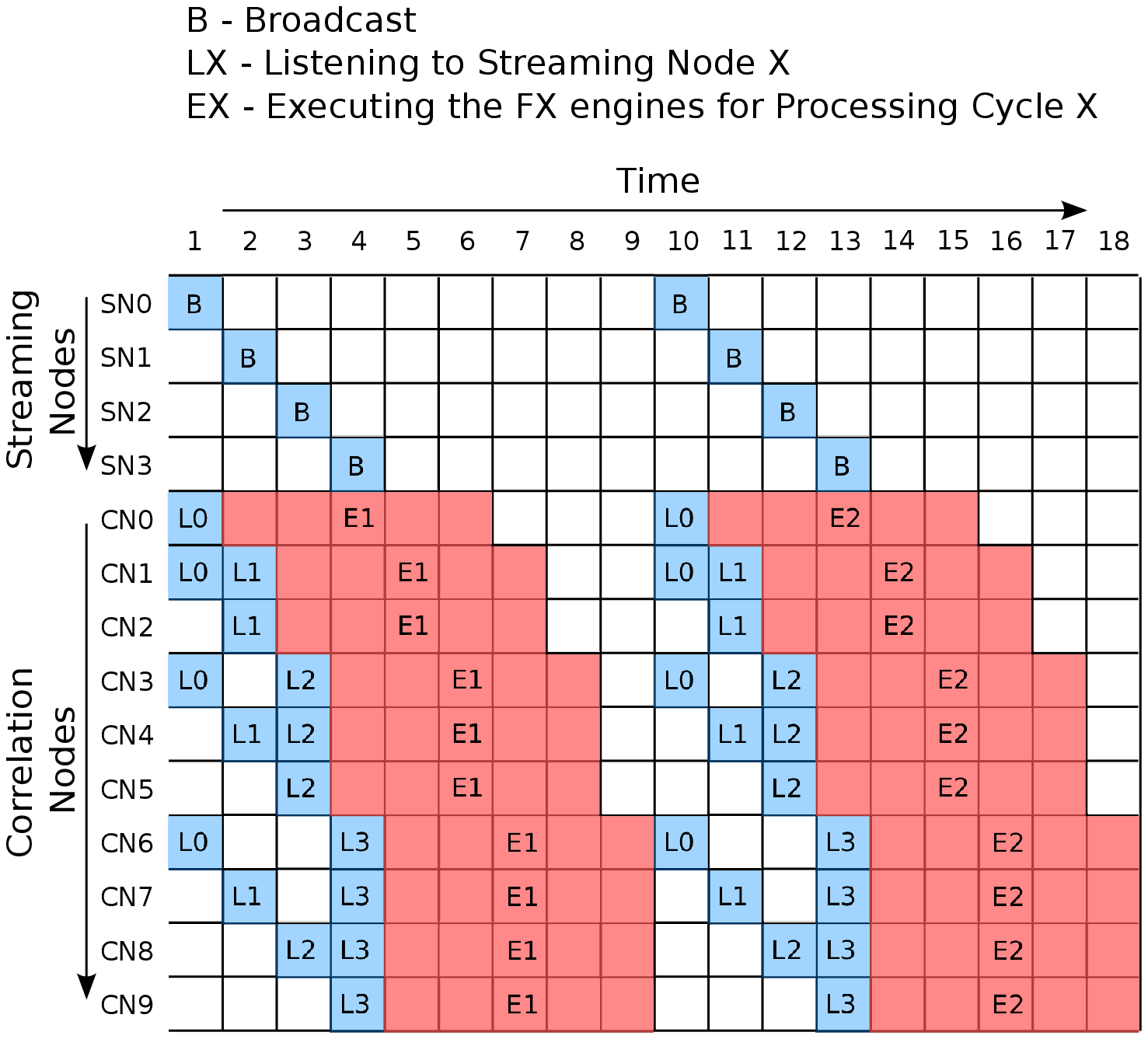}
\label{fig:bdcdiagram1}    
}
\subfigure[Point-to-point]{
\includegraphics[width=0.5\textwidth]{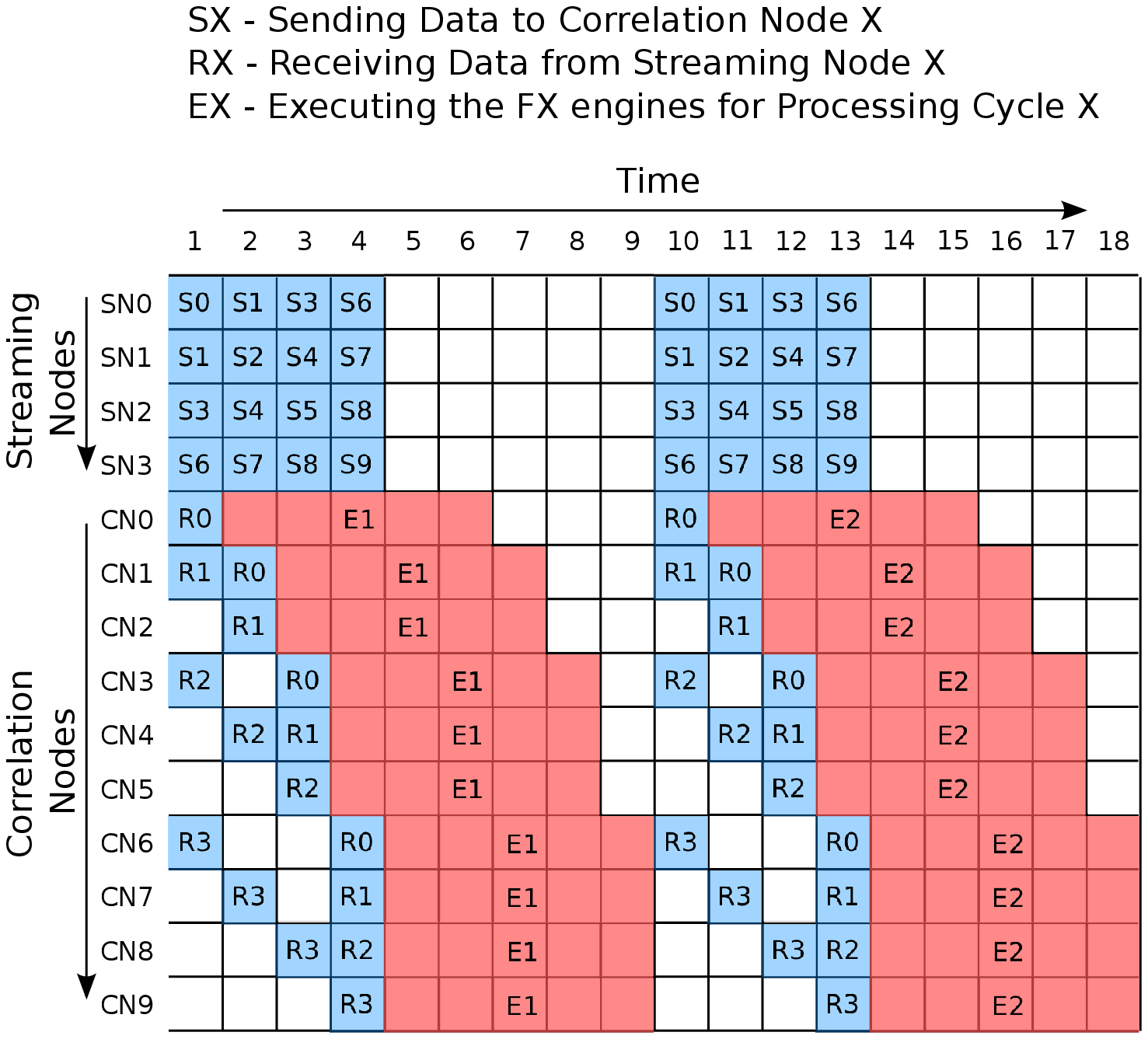}
\label{fig:bdcdiagram2}    
}
\caption{Shown are diagrams of all streaming nodes and correlation nodes with the broadcast routine and the point-to-point routine based on an instance having 10 correlation nodes and 4 streaming nodes. Time is presented in a relative unit where we define a send and receive pair take one unit. In order to simplify the illustration, it is assumed that an execution of the FX engines takes 5 units, while in practice the data transfer to execution ratio varies in a wide range with different configurations. It is also assumed that the broadcast routine works ideally and takes exactly the same as a send and receive pair, while in reality a broadcast usually takes longer depending on implementations.}
\label{fig:bdcdiagram}
\end{figure}

Given the data distributing pattern, there are two methods for the data transfer. One of them is to use a native broadcast routine, which could be either at the MPI level, or a hardware multicast, while the other is to send and receive data in loops using basic point-to-point communications. Figure \ref{fig:bdcdiagram} shows the diagrams of both methods based on the instance given in Figure \ref{fig:broadcasting} with 10 correlation nodes and 4 streaming nodes. 
  
An important fact revealed by Figure \ref{fig:bdcdiagram} is that a native broadcast routine, or even a hardware multicast, would not help improve the overall data transfer efficiency for our models if it is implemented with blocking collective calls. This is because blocking broadcasts for each stream cannot occur concurrently, due to the overlaps between destinations of streaming nodes doing broadcast. In this case, streaming nodes have to broadcast in sequence as shown in Figure \ref{fig:bdcdiagram1}, and this results in the same cost, if not more, as basic point-to-point communications in \ref{fig:bdcdiagram2}. Moreover, non-blocking point-to-point communications do not help either, as there is a limitation of bandwidth rather than latency. However, if a non-blocking multicast routine is available, all broadcasts in Figure \ref{fig:bdcdiagram1} can occur in two relative time units in principle, since there are at most two listening events overlapped on every correlation node. 

To examine the timing in more detail, we first assume that all communications are blocking, every correlation node is assigned with constant correlation tasks, and every streaming node deals with data in a constant size. We also consider $t_t$ the time taken by a single data transfer, which could be either a send and receive pair or a broadcast. Given $t_e$ is the time taken by an execution of the FX engines, $n_s$ is the number of streaming nodes, then the time taken by an entire processing cycle, $t_c$, can be obtained using Equation \ref{eq:bdcb} for both cases show in Figure \ref{fig:bdcdiagram}.

\begin{equation}
\label{eq:bdcb}
t_c = n_s t_t + t_e
\end{equation}

This indicates that by using blocking communications, the larger the number of streaming nodes, the more significant influence data transfers have on the overall performance, which leads to bad scalability. On the other hand, if non-blocking point-to-point communications and double buffering are both applied, then

\begin{equation}
\label{eq:bdcnb}
t_c = max(n_s t_t , t_e).
\end{equation}

Improvements are seen in this case but when $n_s$ is so large that $n_s \cdot t_t > t_e$, the data transfer would still become a bottleneck, and the scalability problem still exists. However, if non-blocking multicasts are used in this model, then

\begin{equation}
t_c = max(2t_t , t_e).
\end{equation}

Hence non-blocking multicasts can largely improve the efficiency, and in this case the time taken by a processing cycle is independent of the cluster size, which results in an excellent scalability as well. 

In practice, collective broadcasts usually mean more overhead and synchronization cost. Hence the actual performance would never reach the ideal situation, especially when using blocking routines. The broadcast routine used by OpenMPI as presented by Fagg et al. \cite{broadcast} utilize a variety of software algorithms. However, the performance of these routines would be less than a true hardware multicast. In this work we only used the OpenMPI broadcast routine due to the limitation of the developing platform. We also completed another implementation based on basic point-to-point communications to verify our analysis above. 

\subsection{Passing Model}

\begin{figure}
\includegraphics[width=0.8\textwidth]{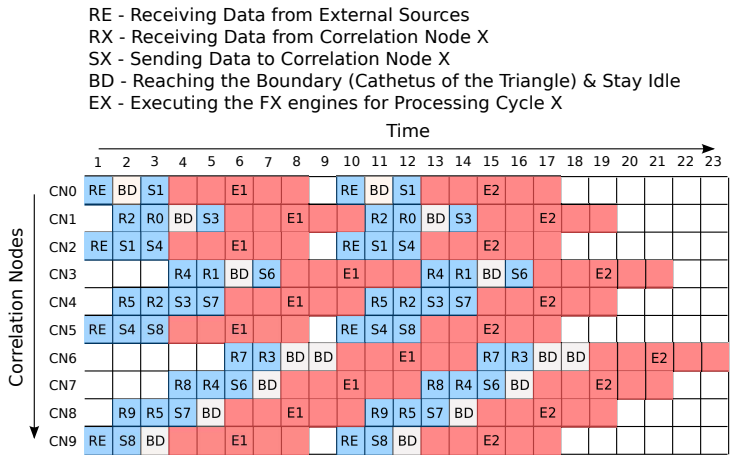}
\caption{Shown is the diagram of the passing model working with 10 correlation nodes. Time is presented in a relative unit where we define a send and receive pair take one unit. In order to simplify the illustration, it is assumed that an execution of the FX engines takes 5 units, while in practice the data transfer to execution ratio varies in a wide range with different configurations.}
\label{fig:psdiagram}    
\end{figure}

In order to avoid the scalability problem while making it suitable for generic environments without requiring a specific non-blocking multicast support, we proposed the passing network topology as our second space-division model. As shown in Figure \ref{fig:passing}, in this model the input data is passed between neighbor nodes. Since all correlation nodes take part in data streaming, dedicated streaming nodes are no longer necessary, which improves the node efficiency as a whole. Figure \ref{fig:psdiagram} illustrates the diagram of the passing model working with 10 correlation nodes, from which we can see that in this model each correlation node deals with four data transfers at most, being two sends and two receives, per processing cycle. With the same definitions as in Equation \ref{eq:bdcb}, the time taken by a processing cycle can be obtained by Equation \ref{eq:psb}.

\begin{equation}
\label{eq:psb}
t_c = 4t_t + t_e
\end{equation}

Similarly, if non-blocking communications and double buffering are applied, then

\begin{equation}
\label{eq:psnb}
t_c = max(4t_t , t_e).
\end{equation}

Thus the data transfer to execution ratio is independent of the cluster scale, which means a better theoretical scalability than the broadcasting model with blocking communication calls. Additionally, by starting data flows from the auto-correlation nodes, it is ensured that every cross-correlation node has an identical distance from the two data sources it claims. As a result, for a cross-correlation node, the two chunks of input data from two different sources arrive at the same time, which saves any extra synchronization cost for input. 

It is noticeable from Figure \ref{fig:psdiagram} that correlation nodes are working asynchronously. More specifically, correlation nodes farther away from data sources have longer delays over time, although a processing cycle on different nodes still costs the same. Taking this into account, given $n_s$ is the number of data sources, which is equal to the number of auto-correlation nodes lying on the hypotenuse of the triangle, and $n_p$ is the number of processing cycles in total, Equation \ref{eq:psb} should be re-written as

\begin{equation}
t_c = \frac{2n_s t_t + (4t_t + t_e)n_p}{n_p}.
\end{equation}

When $n_p \rightarrow \infty$, we have

\begin{equation}
t_c \overset{n_p \rightarrow \infty}{=}  4t_t + t_e.
\end{equation}

Hence the average amount of time taken by a processing cycle is not affected by the delay given the number of processing cycles is sufficiently large. However, the delay of a correlation node, $d$, increases with the distance from the node to data sources, $l$, as given by Equation \ref{eq:delay}, and this could have some negative effects on latency-critical systems.

\begin{equation}
\label{eq:delay}
d = 2l t_t .
\end{equation}

Since the passing model proves to have an excellent scalability in principle, we implemented it in both blocking and non-blocking styles for comparison and analysis. Double buffering is also applied in the non-blocking routine to make all data transfers happen concurrently.

\section{Testing}

Testing was carried out on the Fornax supercomputer, which was designed for data intensive research, especially radio astronomy related data processing. Fornax consists of 96 nodes, each having two Intel Xeon X5650 CPUs, a NVIDIA Tesla C2075 GPU and 72 gigabytes of system memory. The Intel 5520 Chipset is used in the compute node architecture, which enables the NVIDIA Tesla C2075 GPU to work on an x16 PCI-E slot and two 40Gbps QLogic Infiniband IBA 7322 QDR cards on two x8 PCI-E slots. The main storage of Fornax is a 500TB Lustre-based shared file system. One of the two Infiniband networks is dedicated to the communication between compute nodes and the Lustre cluster.

In terms of the software environment, Fornax runs CentOS 6.2 with 2.6.32-131.21.1.el6.x86\_64 Linux kernel. The OpenMPI version adopted in this work is 1.6.0. Default configurations are applied for all communication stacks since our preliminary testing showed that the data transfers almost achieved the theoretical limit of the Infiniband network by doing so. CUDA 4.1.28 library with OpenCL 1.1 support was used for GPU computing. The FFT implementation used for the F-engine was Apple OpenCL FFT 1.6. 

All models presented were tested with the Apple OpenCL FFT for the F-engine and the modified 1xGxG model proposed by Harris et al. \cite{chris} for the X-engine. As the main purpose of this paper is to compare different data distribution models, the F-engine does not include station-based functions other than the FFT, such as fringe rotation. Furthermore, this paper is essentially looking at SKA-scale arrays consisting of 300 to 3000 antennas, and in this case the X engine is more critical as its computational demand scales quadratically with the number of data streams while the F engine scales linearly.  

The metric FLOPS used in all testing results is in single-precision and refers to the actual mathematical operations that are necessary for an FX correlator, which does not include indexing and redundant calculations for optimizing either the GPU memory access or the data transfers. This is a fair method to compare the performance between implementations on different hardware architectures, as the cost of indexing and redundant calculations could vary by several times in order to optimize algorithms for different hardware architectures or different network patterns, while that of the ultimate mathematical operations needed by the correlation algorithm does not change. In our tests all input data is packed in 8-bit integers.

\begin{figure}
\subfigure[\# streams = 128]{
\includegraphics[width=0.5\textwidth]{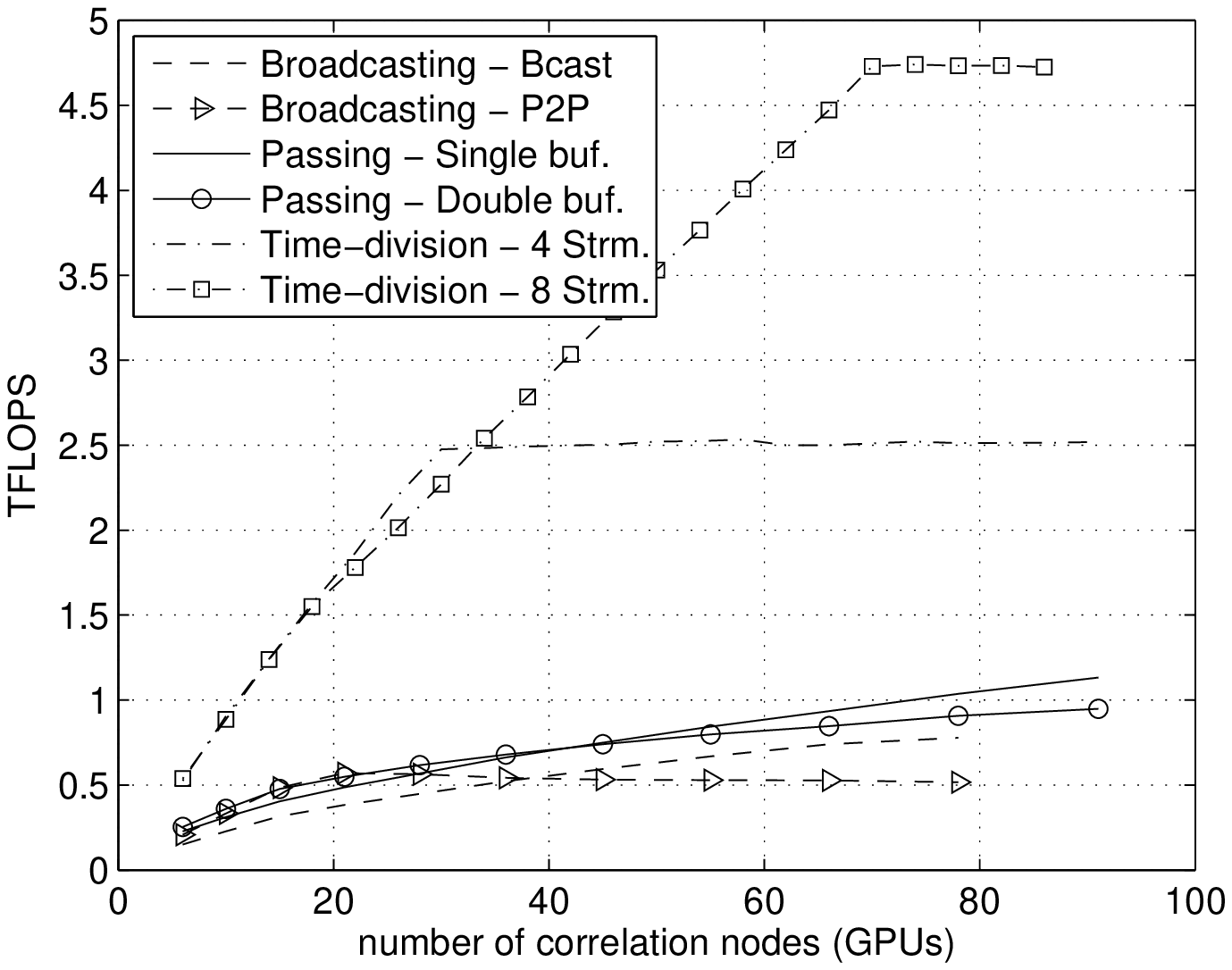}
\label{fig:t128}}
\subfigure[\# streams = 512]{
\includegraphics[width=0.5\textwidth]{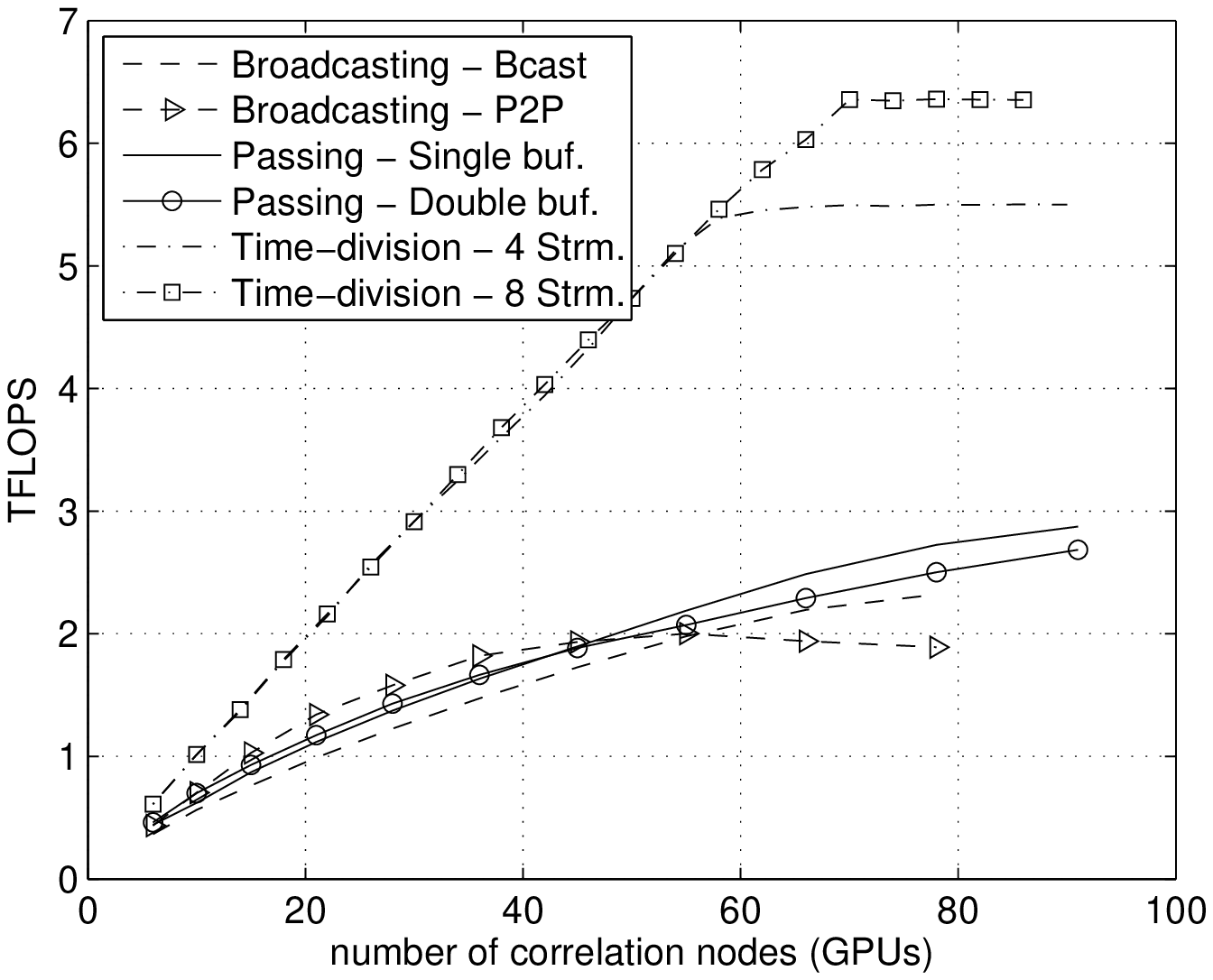}
\label{fig:t512}}
\subfigure[\# streams = 2048]{
\includegraphics[width=0.5\textwidth]{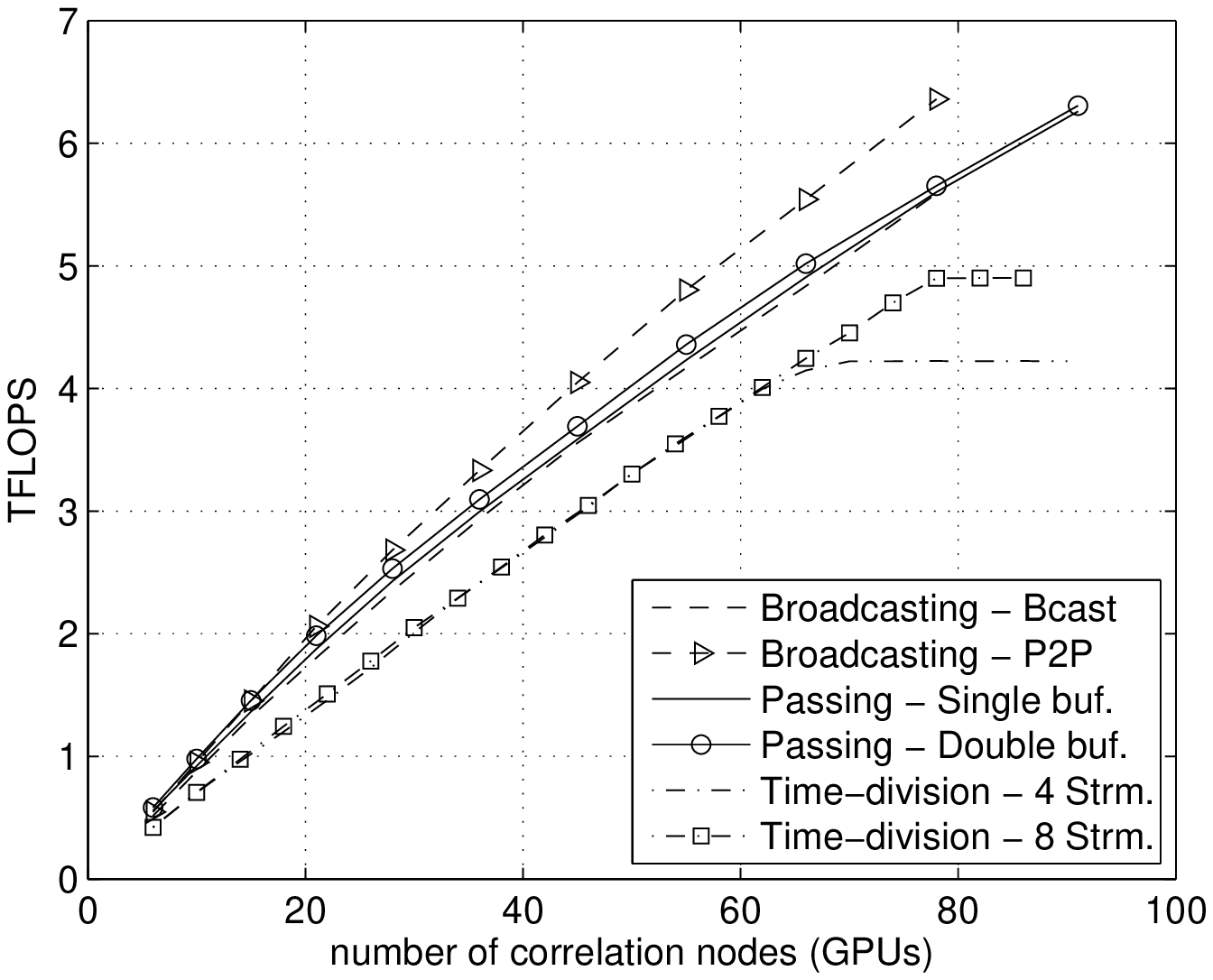}
\label{fig:t2048}}
\subfigure[\# streams = 3072]{
\includegraphics[width=0.5\textwidth]{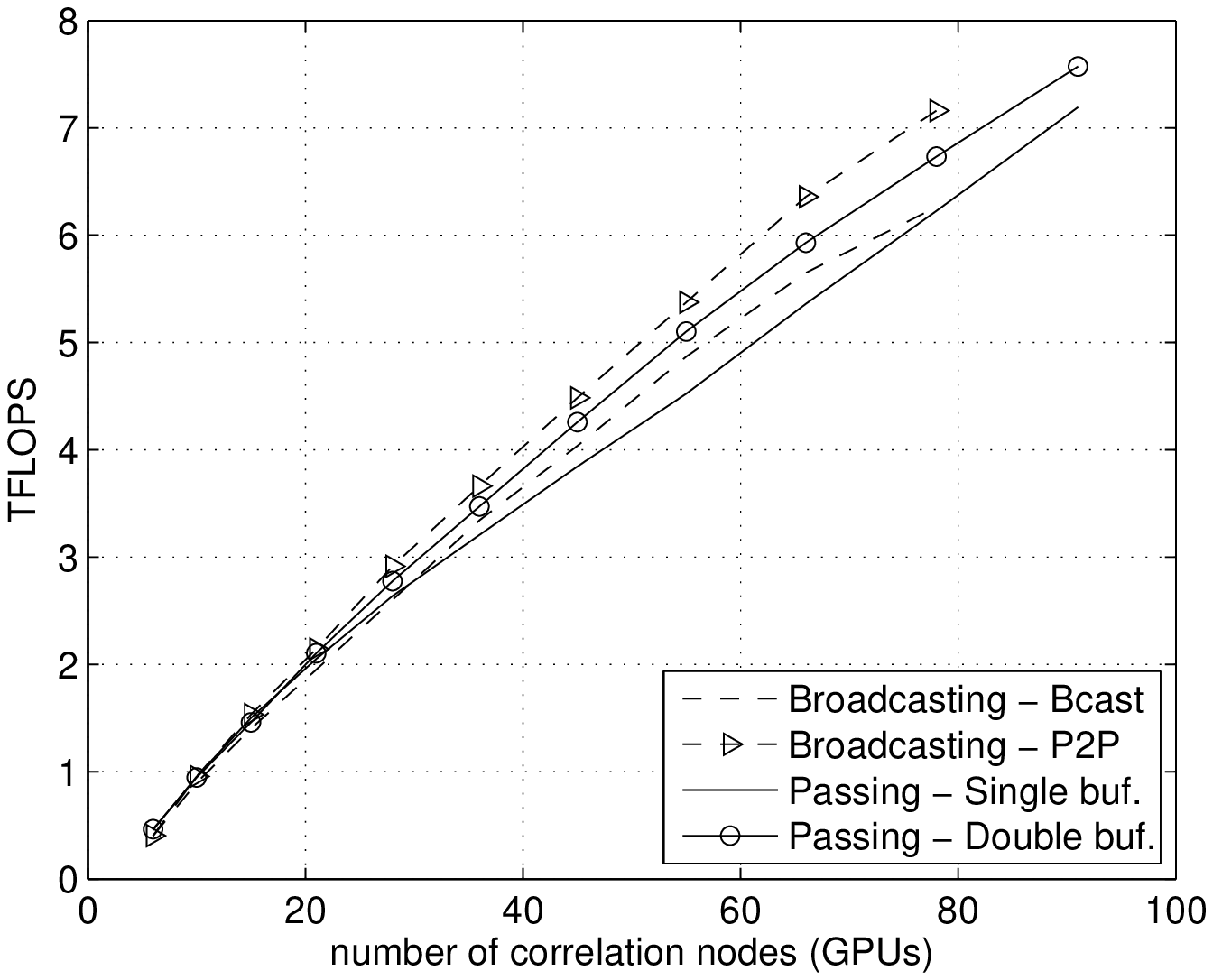}
\label{fig:t3072}}
\caption{Shown is the overall performance in single-precision TFLOPS achieved using various network models and configurations on the Fornax supercomputer. Testing is conducted in four schemes for the number of input data streams. In each scheme, the number of correlation nodes varies from 6 to the maximum number available. The broadcasting model was tested with two subroutines: Bcast which uses the MPI\_Bcast routine; and P2P which uses point-to-point transfer loops. The passing model was tested with single and double buffering for data transfers. The time-division model was tested with 4 and 8 streaming nodes respectively.}
\label{fig:t}
\end{figure}

Testing first investigated how the performance in Tera-FLOPS scales with the number of correlation nodes across all network models. As shown in Figure \ref{fig:t}, testing is conducted in four schemes, with the number of input data streams varying from 128 to 3072. The six configurations used are the broadcasting model with the MPI\_Bcast routine and point-to-point data transfers, the passing model with single buffering and double buffering, and the time-division model with 4 and 8 streaming nodes. When the number of streams reaches 3072, the time-division model is no longer available since a single GPU does not have enough memory to process all the streams. This is not an issue for the space-division models, as they subdivide the problem between GPUs. 

The number of correlation nodes, which is also the number of GPUs executing the FX engines, excluding streaming nodes, varies from 6 to the maximum configuration obtainable on Fornax for each method. An FFT length of 256 is used across all tests, as our preliminary tests showed that the throughput of the Apple OpenCL FFT does not significantly vary with the FFT length in the range from 128 to 2048, and the X stage performance is invariant with respect to FFT length as long as there is sufficient data to feed the massively parallel model of GPU computing.

\begin{figure}
\subfigure[\# streams = 128]{
\includegraphics[width=0.5\textwidth]{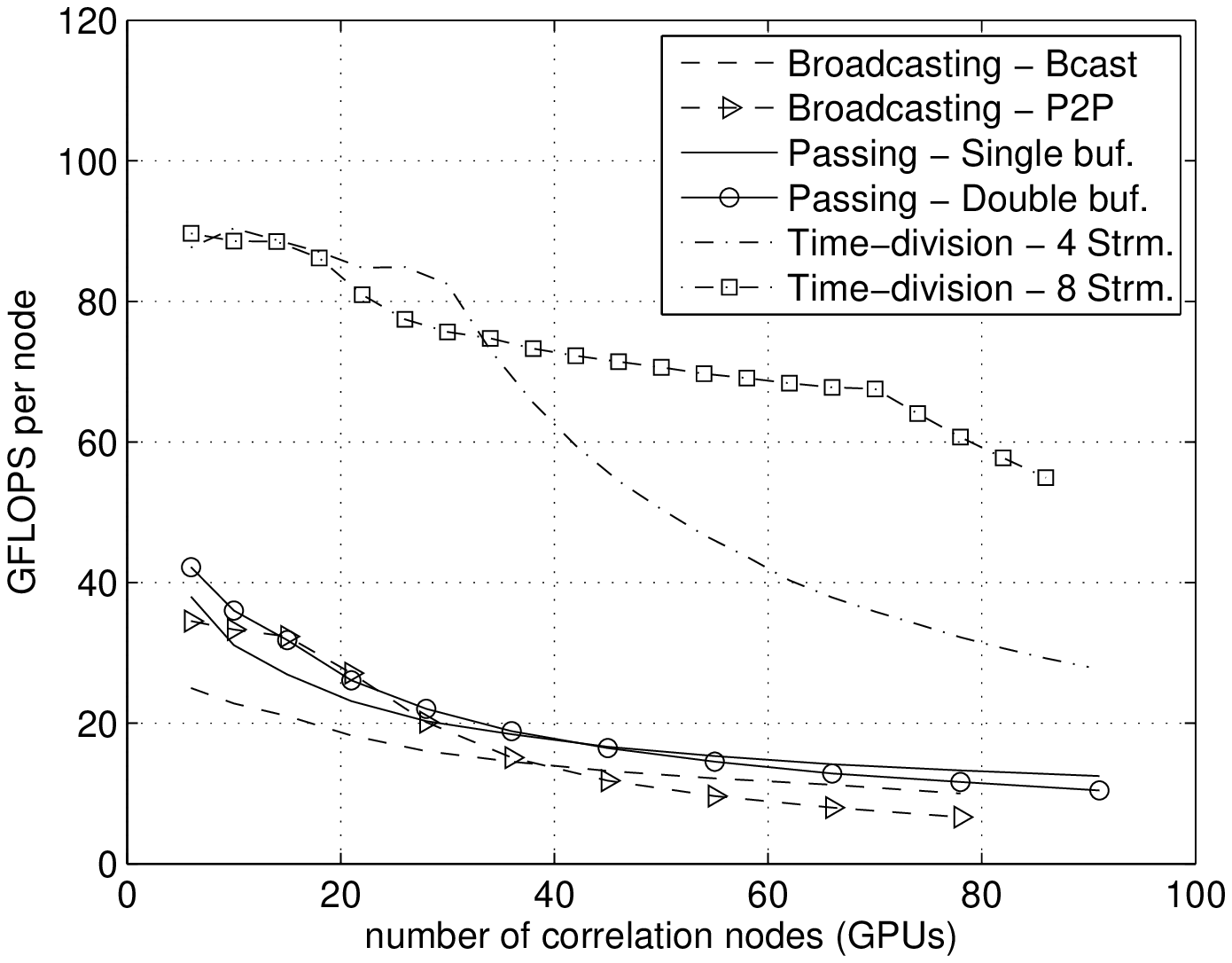}
\label{fig:pn128}}
\subfigure[\# streams = 2048]{
\includegraphics[width=0.5\textwidth]{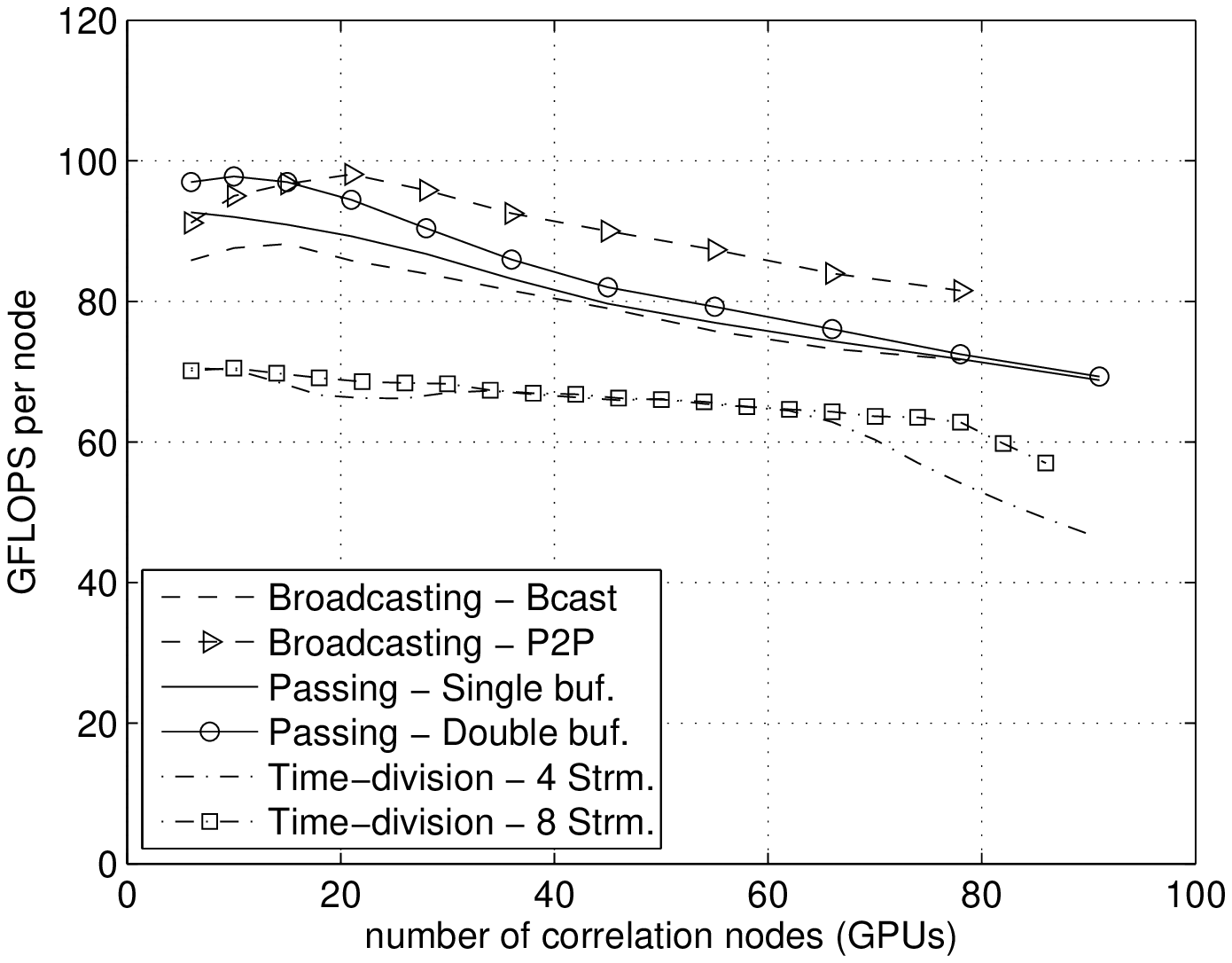}
\label{fig:pn2048}}
\caption{Shown is the performance per GPU in GFLOPS achieved using different network models and configurations. Testing is conducted in two schemes for the number of input data streams. In each scheme, the number of correlation nodes varies from 6 to the maximum number available. The performance shown is averaged over the total number of correlation nodes.}
\label{fig:pn}
\end{figure}

Shown in Figure \ref{fig:pn} is the overall performance averaged over the total number of correlation nodes. This demonstrates the node efficiency across all network models. Based on our preliminary testing, the peak performance that the FX engines achieved on a single GPU is approximately 105 GFLOPS. Thus results shown in Figure \ref{fig:pn} also reveal how the overall performance is affected by the network transport involved for the cluster model. 

\begin{figure}
\subfigure[\# GPUs = 36]{
\includegraphics[width=0.5\textwidth]{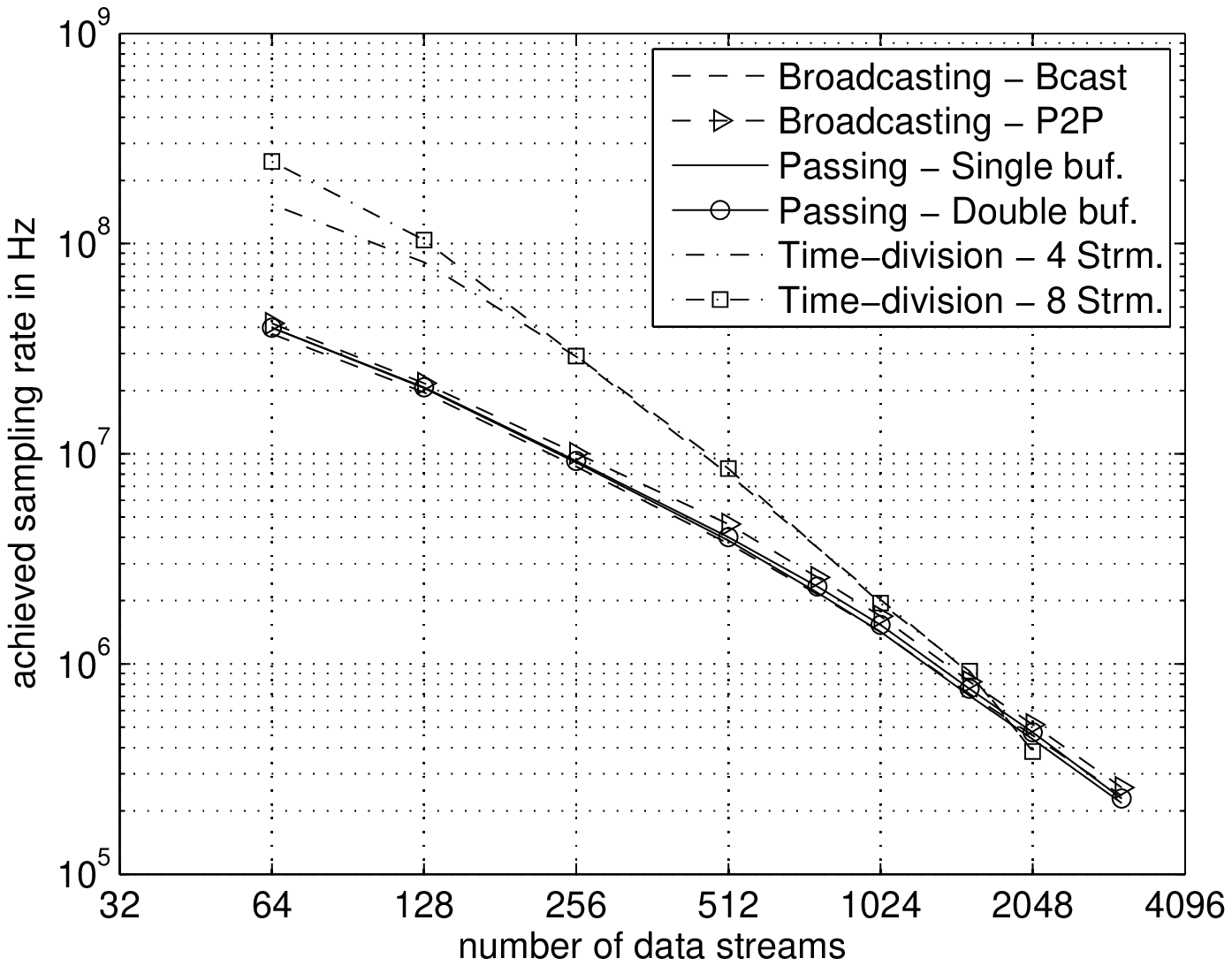}
\label{fig:band36}}
\subfigure[\# GPUs = 66]{
\includegraphics[width=0.5\textwidth]{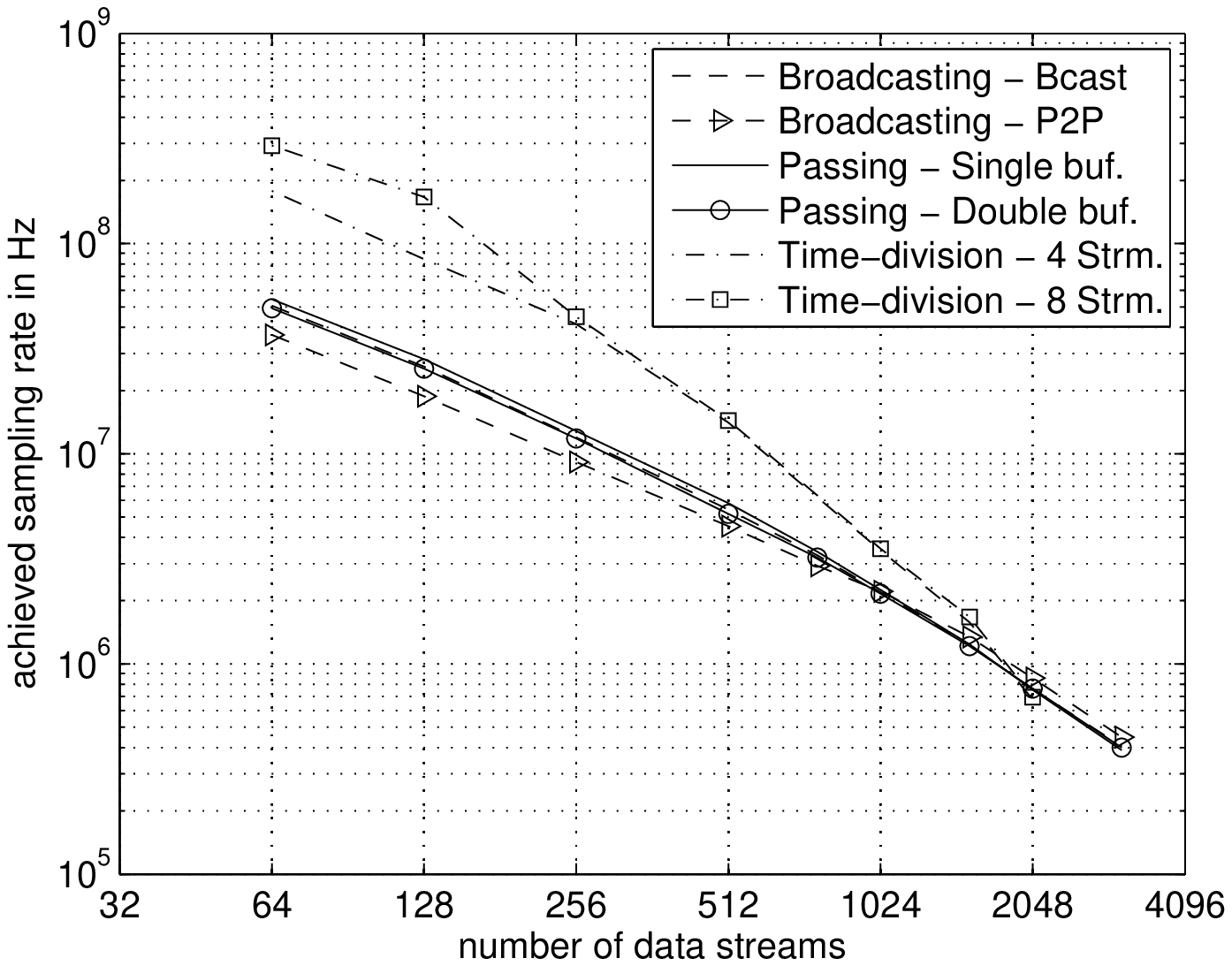}
\label{fig:band66}}
\caption{Shown is the sampling rate achieved using different network models and configurations. Testing is conducted in two schemes varying the number of correlation nodes. In each scheme, the number of input data streams varies from 64 to 3072. Each sample is an 8-bit real integer value.}
\label{fig:band}
\end{figure}

Testing then investigated the sampling rate of input data achieved using our models with different configurations. As shown in Figure \ref{fig:band}, the number of data streams scales from 64 to 3072. The lower limit was chosen as below it the streams are too few to feed the space-division models, while the upper limit was chosen as it is the largest number of data streams likely to be used in the foreseeable future. The time-division model was only tested with up to 2048 data streams due to the limit of the model suitability for the GPU hardware architecture.

The output visibilities were not collected for the performance tests presented above. For correctness tests, we used the Adaptive IO System (ADIOS) devised by Jin et al. \cite{adios} to write visibility files for up to 300 input data streams. By using ADIOS on the Lustre file system, data chunks for different subsets can be filled into a global data space asynchronously, and this enables each correlation node to write visibility data independently while keeping the data in a globally correct order. Taking advantage of the buffering technique and the non-blocking IO mechanism provided by ADIOS, the performance loss caused by IO was too little for us to measure for the testing schemes presented above working with up to 300 input data streams.

\section{Discussion}

Testing results revealed that on current hardware architectures the time-division multiplex model is still the best choice for a GPU cluster correlator when the number of data streams is less than 1024, as shown in Figure \ref{fig:band}. In the range between 1024 and 2048, space-division models start to overtake. Most applications in the foreseeable future will be dealing with less than 1024 streams for which the time-division model is optimal. The only exception we are aware of so far is the SKA Mid Phase 2 which is likely to have 3000 antennas forming a single beam.

However, this is not to say space-division models will be the only way to deal with such large-scale correlation tasks. Rather than the network itself, the GPU architecture is one of the most significant factors leading to the performance turnover between 1024 and 2048 streams. Due to the fact that the output data rate scales quadratically with the number of input streams, the larger the number of input streams is, the bigger proportion of GPU memory the output buffer takes. As the number of input streams increases, at a certain point the output buffer takes so much GPU memory that the input buffer is no longer large enough to hold data that can feed the massively parallel model of GPU computing, which is the major cause of the performance drop. If future GPUs integrate more GPU memory, then it is possible to extend the optimum range of the time-division model. Moreover, the turnover point might shift with a wide range of factors from both hardware and software aspects. This includes but is not limited to the hardware platform and configuration, the optimization of FX engines and the involvement of other correlator functionalities. 

Another implication of the GPU memory limitation is that varying the FFT length would not significantly affect the performance, as long as the GPU memory is enough for both input and output buffers. As the two buffers are both proportional to the FFT length, increasing it does not change the proportion of GPU memory that the input buffer takes, and hence it does not negatively affect the parallel scale of the X-engine. The F-engine might be slightly affected depending on FFT implementations. The performance of the Apple OpenCL FFT we used in this work does not vary significantly for the FFT length up to 4K. However, the FFT length does affect the maximum number of input streams that a single GPU is able to process, as both the input and output buffers need to be at least large enough to hold all intended data at the length of a single FFT. Moreover, this work did not investigate extra-long FFTs beyond 8K, which potentially have some significant impacts on performance. 

\subsection{Network}

The overall throughput that the time-division model can provide depends on the number of streaming nodes. As shown in Figure \ref{fig:t}, the model achieved better scalability when it was given 8 streaming nodes instead of 4. Furthermore, for real-time SKA-scale correlation, 8 streaming nodes are still far from sufficient, otherwise each streaming node needs to handle more than hundreds of gigabytes of data per second, which is orders of magnitude beyond what the current technology can provide. However, unless the throughput is being limited by the streaming nodes, then adding more streaming nodes does not improve performance. Therefore, we did not test the time-division model with more than 8 streaming nodes since this was sufficient in most cases across our testing schemes. 

The time-division model is likely to have excellent scalability even on larger clusters than our testing platform because firstly, based on our testing results shown in Figure \ref{fig:t}, the time-division model achieves a more linear scalability than space-division models, and secondly, the scalability is only limited by the number of streaming nodes rather than the network topology and a series of factors for space-division models. 

It seems from Figure \ref{fig:t} that the broadcasting model achieves better performance than the passing model on the same number of GPUs when the number of data streams is large. However, this is based on the prerequisite that a considerable number of extra nodes are allocated as streaming nodes, as given by Equation \ref{eq:oc_os}.

The passing model is promising in solving large scale correlation problems in the future. An obvious advantage is that it does not need any dedicated streaming nodes to re-organize and distribute data, for the auto-correlation nodes are the only nodes that receive input from external sources, and are able to receive streaming data in its original form. The topology also prevents network bottlenecks to a large extent, as the number of data transfers that each node deals with does not scale. In principle, the performance would scale linearly. Our passing model testing results showed a near linear trend in which the performance falls behind the broadcasting model when large datasets are applied because the passing model does not perfectly suit the switch-based topology applied on Fornax. It is likely to scale better on supercomputers with multi-dimensional torus topology, in which neighbor nodes have dedicated communication channels, as well as clusters with custom networks built-in to match our model.

\subsection{FX Engines}

For the space-division models, it is debatable whether or not to process FFT and CMAC coherently on the same node. Redundant FFTs are introduced if they are on the same node, as correlation nodes in the same row or column claim the same input data streams. On the other hand, if they are processed on separate nodes, the network load would increase by several times, as the data for a sample is usually packed in a small number of bits before FFT and is expanded to the complex floating-point format in 64 bits afterward. Thus it is ultimately a trade-off between compute and network. From Figure \ref{fig:pn} we can see that even using our optimum network model under its favorable configuration, the performance per node is still reduced by approximately 30\% when the number of total correlation nodes scales up to 90, compared with the peak single GPU performance which is 105 GFLOPS. This indicates that for large-scale correlation problems, the network is where bottlenecks would mostly appear, rather than compute. Additionally, in large-scale correlation systems, the FFT only takes a small proportion of the entire FX correlator in terms of execution time, due to the fact that the computational demands of the FFT scales linearly with the number of data streams while that of the CMAC scales quadratically. In this case, redundant FFTs introducing minor performance loss are more desirable than increasing the network load by potentially an order of magnitude. 

The GPU FX engines used in this work achieved approximately 10\% of the capacity of C2075 GPUs by using the metric that counts only the mathematical operations. To include the indexing and redundant calculations, a factor ranging from approximately 2 to 4, depending on the FFT length, the accumulation size and the network model, needs to be multiplied. The optimization techniques in GPU computing change significantly with hardware architectures, while the FX engines used in this work was designed several generations ago in terms of the GPU architecture. It is likely that FX engines optimized for newer GPU architectures can largely improve the performance of the GPU cluster correlator. The network models presented in this paper are applicable to such newer GPU FX engines, and also can be integrated in systems based on other hardware architectures such as CPUs and FPGAs. 

\subsection{Visibility Data}

In our tests, the output visibility data was only written to files on the Lustre file system for up to 300 input streams for correctness verification. There are two reasons behind this, firstly for next-generation telescopes which generate enormous amount of data, the visibility data is not likely to be stored on hard disks, but rather being streamed directly to post-processing stages. In this case, the visibility data resulted from our models would need to be re-ordered in time-stamped sub-bands or potentially other patterns. This re-formatting process can occur concurrently with the GPU X-engine, and be fully parallelized and completed on each correlation node independently for the time-division model, where every correlation node processes all baselines. 

For the space-division models, visibility data containing a subset of the baselines on each correlation node can be first split into sub-bands locally, and then gathered on post-processing nodes for all baselines. Each post-processing node deals with a sub-band, so that the corner turning and imaging algorithms can be applied concurrently on each node without gathering all data onto a single node. There needs to be a streaming many-to-many network connecting correlation nodes and post-processing nodes, which is similar to what we implemented in our models to send input data from streaming nodes to correlation nodes.

Secondly, the output data rate is usually less critical than the input. Shown in Table \ref{tab:ratio} are calculated input and output data rates for the SKA Phase 1 correlator from Ford et al. \cite{SKA}. As seen in these figures, the output data rates are much lower than the input. Hence in immediate future the actual problem we are likely to face is still an input-limited correlation system rather than output-limited. Correlators for the full scale SKA might be output-limited. However, while trying to meet science requirements, the final design will also largely depend on how relevant technologies develop in the next decade and how the cost can be controlled in a reasonable range. Taking these into account the implementation of the output network is left to future work.

\begin{table}
\caption{Shown is a summary of projected total input and output data rates for SKA Phase 1 correlator from Ford et al. \cite{SKA}. The data rates are given for both low and mid frequency SKA 1. The reader is referred to Ford et al. \cite{SKA} for further details on the calculation of these values.}
\label{tab:ratio}       
\begin{tabular}{c|cc}
\hline\noalign{\smallskip}
Case & SKA 1 Low & SKA 1 Mid  \\ \hline
Total Input Data Rate & 18.24 TB/s & 750 GB/s \\
Total Output Data Rate & 2.405 TB/s & 136.7 GB/s\\
\noalign{\smallskip}\hline
\end{tabular}
\end{table}

\subsection{Real-time Capacity}

Next-generation telescopes are likely to have an entirely streaming work flow in order to reduce the expense of storing intermediate data. This requires all correlation data to be processed in real time. However, being limited by current technology, when the problem size approaches the SKA scale, the sampling rate of input signals achieved in our testing, as shown in Figure \ref{fig:band}, falls far behind what is required. This situation can be changed in three aspects in the future. Firstly, new generation GPUs are likely to double the performance every other year, and by the time SKA-scale telescopes come into reality, the newest GPUs would be at least an order of magnitude faster than what we used in our testing. 

Secondly, as the GPU computing industry grows and more developer resources become available, optimizing the GPU FX engines would become easier. The hardware architecture and compiler would also evolve towards a direction that provides simpler ways to utilize more of the GPU capacity. While the GPU FX engines used in this work still have considerable space to optimize on current GPU architecture, evolving with new technology becomes even more important. 

Thirdly, for SKA-scale real-time correlation, it would eventually be necessary to scale our models on much larger clusters. This could be at the level of 10 to 100 times as large as our testing platform. In this case the network would become increasingly critical, and implementing our passing model on a cluster with multi-dimensional torus network would be a promising solution. The time-division model is another choice if future GPU architectures allow a single GPU to process all baselines of the telescope array. 

\section{Conclusion and Future Work}

This work has investigated several ways to scale a single GPU based software correlator to clusters. We have investigated two major strategies, which are the time-division and space-division multiplex systems, and compared the performance over a range that is large enough to meet the requirements in the foreseeable future. Our testing results have shown that for numbers of data streams smaller than 1024, the time-division model is more efficient, while the passing topology of the space-division model showed advantages for large numbers of streams due to the more efficient use of the GPU memory.

As it is difficult to predict the development of technology in the next decade, it is still too early to make statements as to how achievable it is to build a real-time GPU cluster correlator for a 3000-antenna telescope such as the SKA Mid Phase 2. Meanwhile there is still considerable space for our models to be optimized. Future work will therefore firstly focus on replacing the GPU FX engines with newer and more optimized implementations, the xGPU developed by Clark et al. \cite{xGPU} for instance, and optimizing models for real world projects. 

In terms of the network patterns, there is a possibility of designing a hybrid model combining advantages of both space-division and time-division models. Orthogonal correlation triangles separating frequency channels prior to the CMAC stage is another promising direction to investigate, which can generate output visibilities in a more friendly pattern for post-processing but involves the design of a complex communication network between FFT and CMAC. Some non-performance-critical functionalities such as the delay compensation are also to be added to make a fully integrated system, as this work only investigates the compute intensive stages of an FX correlator. With the high flexibility of a software correlator, it is also sensible to integrate other functional techniques into the correlation flow, such as a coherent fast transient detector proposed by Law et al. \cite{transient} which needs to be placed between conjugate multiplications and accumulations within the X-engine.

\begin{acknowledgements}
The work was supported by iVEC through the use of advanced computing resources located at iVEC@Murdoch and iVEC@UWA.
\end{acknowledgements}


\begin{thebibliography}{}
%
%



\bibitem{DiFX}
A.~T.~Deller, S.~J.~Tingay, M.~Bailes and C.~West.: 
DiFX: A software correlator for very long baseline interferometry using multiprocessor computing environments. 
Publications of the Astronomical Society of the Pacific, 119, 318--336 (2007).

\bibitem{DiFX2}
A. T. Deller, W. F. Brisken, C. J. Phillips, J. Morgan, W. Alef, R. Cappallo, E. Middelberg, J. Romney, H. Rottmann, S. J. Tingay and R. Wayth.: 
DiFX2: A more flexible, efficient, robust and powerful software correlator.
Publications of the Astronomical Society of the Pacific, 123, 275--287 (2011).


\bibitem{richard}
R. Dodson et al.:
Astronomical HPC: Intel's Many Integrated Core in Astronomical Applications.
Intel Supercomputing Conference (2012).


\bibitem{lofar}
J.~W. Romein, P.~C. Broekema, E.~Meijeren, K.~Schaaf, and W.~H. Zwart.:
Astronomical real-time streaming signal processing on a Blue Gene/L supercomputer.
Proceedings of the eighteenth annual ACM symposium on Parallelism in algorithms and architectures, 59--66 (2006).


\bibitem{lofarcorr}
J. W. Romein, P. C. Broekema, 	J. D. Mol and R. V. Nieuwpoort.:
The LOFAR correlator: implementation and performance analysis.
Proceedings of the 15th ACM SIGPLAN Symposium on Principles and Practice of Parallel Programming, 169--178 (2010).


\bibitem{IBM}
R.~V. Nieuwpoort and J.~W. Romein.:
Correlating radio astronomy signals with many-core hardware.
International Journal of Parallel Programming, 39(1), 88--114 (2011).


\bibitem{cots}
K. Schaaf and R. Overeem.:
COTS correlator platform.
Experimental Astronomy, 17, 287--297 (2004).

\bibitem{chris}
C.~Harris, K.~Haines, and L.~Staveley-Smith.: 
GPU accelerated radio astronomy signal convolution.
Experimental Astronomy, 22, 129--141 (2008).

\bibitem{mwa}
R.~B. Wayth, L.~J. Greenhill, and F.~H. Briggs.:
A GPU based real-time software correlation system for the Murchison Widefield Array prototype.
Publications of the Astronomical Society of the Pacific, 121(882), 857--865 (2009).

\bibitem{xGPU}
M. A. Clark, P. C. La Plante and L. J. Greenhill.:
Accelerating radio astronomy cross-correlation with Graphics Processing Units.
arXiv:1107.4264v2 (2012).

\bibitem{SKA}
D. Ford and A. Faulkner and P. Alexander.:
A Software Correlator for SKA.
The SKA Publications, Memo 139 (2012).


\bibitem{broadcast}
G. E. Fagg and J. Pjesivac-grbovic and G. Bosilca and J. J. Dongarra and E. Jeannot.:
Flexible collective communication tuning architecture applied to open MPI.
2006 Euro PVM/MPI (2006).


\bibitem{adios}
C. Jin, S. Klasky, S. Hodson, W. Yu, J. Lofstead, H. Abbasi, K. Schwan, M. Wolf, W. Liao, A. Choudhary, M. Parashar, C. Docan, and R. Oldfield.:
Adaptive IO System (ADIOS). 
In Cray User Group (2008).

\bibitem{transient}
C. J. Law and G. C. Bower.:
All Transients, All the Time: Real-Time Radio Transient Detection with Interferometric Closure Quantities.
arXiv:1112.0308v2 (2012).


\end{thebibliography}


\end{document}